# Comparing architectures of mobile applications

Krešimir Fertalj[1], Marko Horvat[2]
[1]*Faculty of Electrical Engineering and Computing, University of Zagreb*
*Unska 3, HR-10000 Zagreb, Croatia*
[2]*Croatian Railways Ltd.*
Branimirova 9a, HR-10000 Zagreb, Croatia
*E-mail:* [kresimir.fertalj@fer.hr](kresimir.fertalj@fer.hr), [marko.horvat@hznet.hr](marko.horvat@hznet.hr)

**Abstract.** *This article describes various advantages and disadvantages of SMS, WAP, J2ME and Windows CE technologies in designing mobile applications. In defining the architecture of any software application it is important to get the best trade-off between platform's possibilities and design requirements. Achieving optimum software design is even more important with mobile applications where all computer resources are limited. Therefore, it is important to have a comparative analysis of all relevant contemporary approaches in designing mobile applications. As always, the choice between these technologies is determined by application requirements and system capabilities.*

**Keywords:** *SMS, MMS, WAP, J2ME, Pocket PC, Windows CE, mobile devices, PDA, mobile architectures, mobile computing.*

## 1. Introduction

Although personal computers are regularly becoming more and more powerful with greater storage, processing and data presentation capabilities, the art of meaningful software design hasn't died-off. Quite the contrary, it is alive and well as always – especially in the emerging field of mobile applications. Mobile devices, such as mobile phones, pocket computers, smart wrist watches and other similar devices, have very limited resources. This is perhaps most evident when we compare their capabilities to the capabilities of non-mobile or "regular" platforms like desktop computers. Mobile devices are at a disadvantage due to restrictions in their size and power consumption. These restrictions are defined by current technology but also by shear practicality. Because of their nature, mobile devices have upper and lower size limitations. Their usage also has to be uncomplicated and fast. It is unreasonable to assume that people will ever prefer to carry around large, heavy or cumbersome devices compared to small, light and sleek gadgets. Roughly speaking, features of mobile devices can be divided into three categories. First of all, it is unlikely that the size of mobile device screens can significantly change, at least in the next foreseeable period. On the other hand, graphics and sound quality improves regularly. For example, it might be expected that text-to-speech algorithms will be implemented in hardware or software of future mobile phones. Secondly, processor power, memory storage and battery durability are bottlenecks but they improve with every next generation of mobile devices. Third feature of mobile devices is connectivity. Nowadays GPRS (Class 10) [1] is virtually commonplace as well as EDGE [2] and 3G/4G mobile phones [3], [4] with significantly larger bandwidths. IrDA (infrared) [5], [6], [7] interface is a feature of almost all business-class mobile devices. Bluetooth [8], [9] is also becoming more common. Some "exotic" models of mobile devices have FM radio and TV tuners. Wi-Fi (WLAN) [10], [11] is probably the next connectivity barrier that will be conquered. The conclusion from all of this is that today's mobile devices, mobile phones in particular, have enough features that allow software developers to create useful applications for them. Only about a year ago J2ME-enabled [16] and Windows CE-enabled [18] devices were rare but today there are plenty of programmable mobile devices and therefore plenty of incentive to make useful mobile computing solutions.

## 2. Classification of mobile architectures

Mobile applications and their architectures can be classified in several different ways. The most significant classifications are by application applicability, architecture, functionality and range.

## 3. Applicability

Applicability of a mobile application can be platform-specific or platform-generic. In other words, a mobile application can be applicable only on a specific mobile device, or the application can be independent of the mobile platform type. Depending on the target user group it is possible to design applications for a single, specific mobile device or, if the user group is broad and undefined, it is important to make applications work on a variety of mobile devices. If a mobile application is platform-specific developers can extract maximum features from the same volume of code by directly using the API of the specific device but the drawback is that all other users who don't have the targeted device will not be able to run this mobile application. If users, for example, will be from one company only and will have the same mobile phone or handheld device this isn't a big problem, but it poses a significant drawback if the mobile application has to be used by the general public – the widest possible range of users. In this case it is necessary to make mobile applications that will work if not on all than on all significant mobile platforms. A second possibility that is especially exploited by developers of J2ME games is to make several versions of the same application where each version works on only one type of mobile device. In this case user must find and install the version that was developed only for his mobile phone or handheld device. Perhaps the best solution would be to develop one generic less-than-optimal application that works on all mobile devices and a number of optimized applications for individual devices. However, it is questionable that the generic application can have all the features of the platform-specific application. In any case, there is no single perfect solution for all cases and the choice of mobile application design is an important decision in the development process.

|  | Generic | Specific |
|---|---|---|
| SMS | ● |  |
| WAP | ● |  |
| J2ME | ● | ● |
| Windows CE | ● | ● |

**Table 1 Applicability constraints**

As can be seen in Table 1, J2ME and Windows CE applications can use generic functions or, in order to enhance functionality, they can use specific API functions that are specific to certain devices such as Nokia, Motorola, Siemens, Psion, etc. J2ME classes provided by mobile phone manufactures are called OEM-Extensions. These classes extend standard J2ME functions with functionalities to Bluetooth and IrDA hardware, better rendering of graphics and animation, restricted access to voice and SMS communications, etc.

## 4. Architecture

The second paradigm for mobile application classification is architecture. This is illustrated in Table 2. SMS (and MMS) applications always have complex client/server architecture [12] with several layers of business logic on the server. As shown in
Figure 1, these applications require a heavy multi-layered telecommunication infrastructure and close cooperation between application developer and Telecom. For example, typical SMS application would use several different databases, modules written in C, C++, Java or .NET that all communicate by CORBA, XML/SOAP or just with sockets.

|  | Client/Server | Single layer |
|---|---|---|
| SMS | ● |  |
| WAP | ● | ● |
| J2ME | ● | ● |
| Windows CE | ● | ● |

**Table 2 Architecture classification**

SMS and MMS applications cannot be developed without a contract with a Telecom and permission to use Telecom's infrastructure such as IN and SMSC.

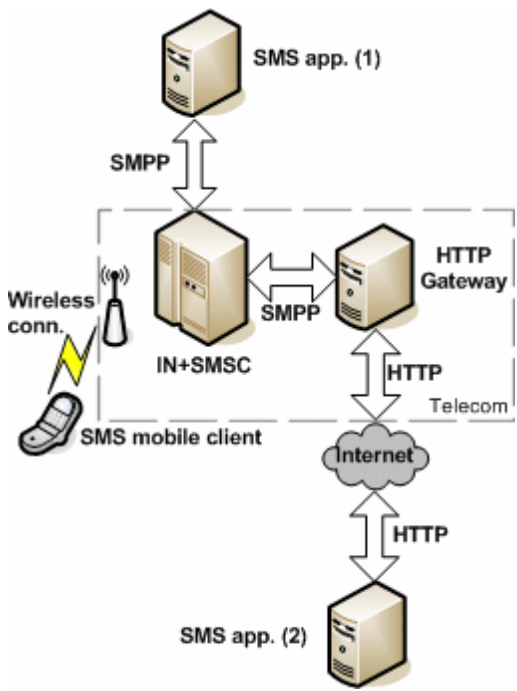

**Figure 1 SMS application architecture**

WAP [13], [14] is a markup language similar to HTML – WAP pages are very similar to standard Web pages. WAP applications consist of a number of WAP pages that can exist on any Internet Web server as shown in Figure 2. WAP pages are relatively lightweight and accessible by (virtually) all mobile phones on the market today. J2ME and Windows CE compared to other two types of mobile applications are really "desktop-like".

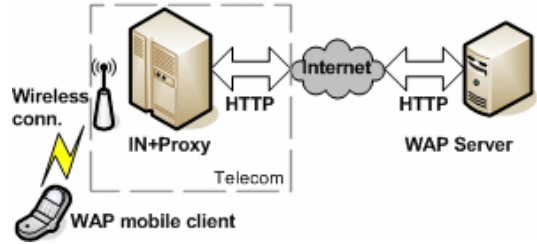

**Figure 2 WAP architecture**

They are specific, constrained, developed and run on a mobile device but they are essentially identical to applications found on desktop computers. The enabling technology with J2ME and Windows CE is the ability to wirelessly connect a mobile application to the Internet and wirelessly send and receive data. Such mobile application can be a client in a client/server multi-tier system, or it can be a standalone application without a connection to the server. J2ME and Windows CE provide the best platform for the implementation of elaborate business logic on the client, e.g. mobile device.

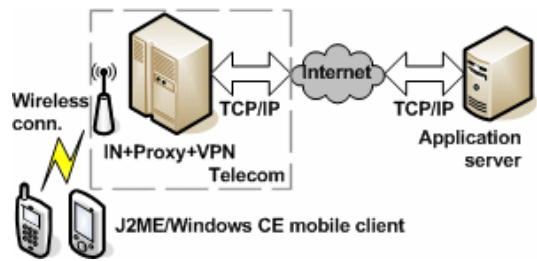

**Figure 3 Architecture of J2ME and Windows CE applications**

## 5. Functionality

Functionality is the third paradigm for the classification of mobile applications and it is closely related to the architecture. As illustrated in Table 3, mobile application functionality can be divided in three distinct categories: "messaging systems", "Web-based" and "desktop-like" applications. Functionality also implies the ability of mobile applications to implement a comprehensive user interface. SMS applications are in essence a messaging system where user and system exchange textual information - they support only textual user interfaces. The only difference between SMS and MMS applications, in this respect, is that MMS applications along with textual information can also supply images, animations and sounds.

|  | Msg. Systems | Web-based | Desktop-like |
|---|---|---|---|
| SMS | ● |  |  |
| WAP |  | ● |  |
| J2ME |  |  | ● |
| Windows CE |  |  | ● |

**Table 3 Functionality classification**

|  | Text-only | Simple | Complex |
|---|---|---|---|
| SMS | ● |  |  |
| WAP |  | ● |  |
| J2ME |  |  | ● |
| Windows CE |  |  | ● |

**Table 4 User interface capability**

A server parses text messages (i.e. SMS) from the user (Figure 4) and extracts valuable information, e.g. vehicle's license plate, streetcar's route, highway or railway route, unique code of TV or theatre show, movie in cinema or some product seen on a TV marketing show.

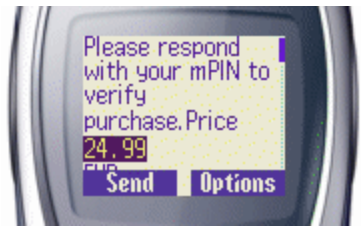

**Figure 4 The user interface of an SMS application**

The possibilities for SMS applications are quite endless. Examples are paying parking tickets, buying groceries, making cinema reservations, listing ferry schedules, etc. WAP can be thought of as a "stripped-down Web". They have more capable interface with text, different fonts, 2-bit colored bitmaps, textboxes, option buttons and hyperlinks. WAP pages can be static or they can be a presentation layer of a multi-tier system and, therefore, dynamic. Their appearance, as shown in Figure 5, is simple and rather rudimentary [15].

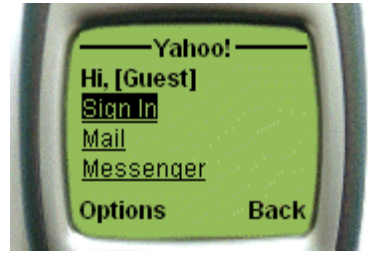

**Figure 5 Appearance of WAP applications**

On the other hand, J2ME and Windows CE provide the best user interface that is most similar to desktop systems. Their appearance, capabilities and development methods are virtually identical to non-mobile applications. J2ME applications [17] support keyboard events, text boxes and labels, buttons, radio buttons, lists, date and calendar controls, marquees, progress bars, images and message boxes. All are optimized for small mobile phone screens. J2ME applications have pretty elaborated graphics and multi-threading support so they are often used for development of mobile games, as can be seen in Figure 6.

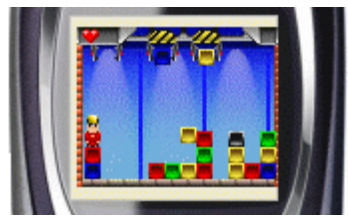

**Figure 6 Typical J2ME game**

Development tools such as Sun's Forte, Borland's JBuilder and Eclipse represent powerful environments for the development of personal and enterprise J2ME applications. Many development tools also provide their own software emulators of J2ME-enabled mobile devices. Companies such as Nokia and Siemens allow J2ME developers free download of their own mobile phones' emulators. In contrast to J2ME-enabled devices, Windows CE handheld computers have larger, color touch-screens. This enables them to have a "desktop-like" user interface [18], [19]. In addition to all J2ME controls Windows CE devices support dropdown lists, context menus, tree views, list views, list boxes, image lists, data grids, open/save file dialogs and several other useful controls. They also support handwriting recognition and portable SQL databases (Microsoft SQL Server CE).

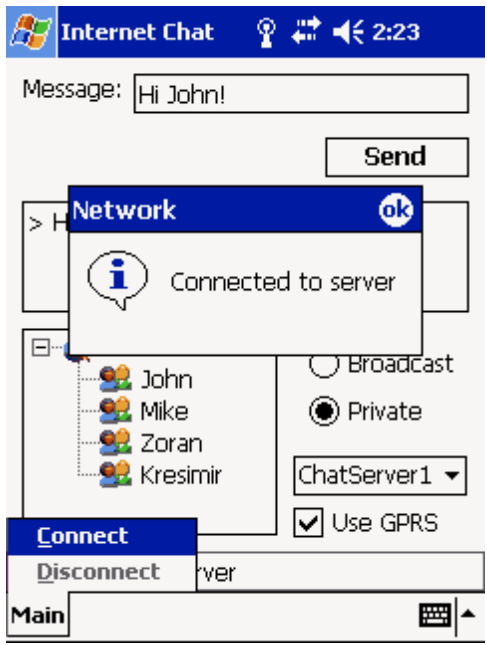

**Figure 7 Typical Windows CE application**

Current Windows CE version is called Windows Mobile 2003. It's build around Windows CE .NET 4.2 operating system. Windows Mobile 2003 has preinstalled .NET-based compact framework (".NET CF"). Development for this platform is possible in C#.NET and VB.NET languages that produce managed code, i.e. code that runs on a .NET virtual machine. It is important to notice that Windows CE is a real-time operating system and has faster code execution than J2ME. Also, Windows CE platform has full multithreading support. The development tool for Pocket PC devices, e.g. Windows CE operating system, is Microsoft Visual Studio. This is a versatile and capable development tool with auto-deployment of executable code and its components, remote debugging support, series of emulators, etc. In order to make multi-tier J2ME and Windows CE applications possible the mobile device must have wireless connection by CDMA, GSM/GRPS/EDGE or WLAN. J2ME and Windows CE have classes and methods that support a wide range of communication protocols including TCP/IP, UDP, HTTP and sockets. In case of J2ME careful attention must be paid to the version of MIDP implementation on the mobile device. The safest thing to do is to use only the HTTP protocol on default socket 80 since it is defined in MIDP 1.0 and, therefore, available on all J2ME-capable mobile devices. The newest MIDP 2.0 is far more powerful and it additionally defines standard programming interfaces for voice calls, access to address book, inbox, Bluetooth, IrDA and many other mobile phone features.

## 6. Range

Table 5 displays range or spread of different mobile technologies today.

|  | Max | Med. | Min |
|---|---|---|---|
| SMS | ● |  |  |
| WAP | ● |  |  |
| J2ME |  | ● |  |
| Windows CE |  |  | ● |

**Table 5 Range**

As can be expected SMS is the most present mobile technology and represents the most interesting technology for omnipresent mobile applications. The same goes for WAP too; WAP 1.1 and WAP 2.0 standards are features of all modern mobile phones. J2ME is also a very widespread mobile technology. Manufacturers do not longer consider a hardware implementation of J2ME virtual machine as an advanced or costly feature. In the last year and a half J2ME is a commonplace feature of almost all middle and higher-class mobile phones from Nokia, Motorola, Siemens, Samsung and others. Spread of Windows CE devices is not as high as the spread of J2ME mobile phones but one can certainly expect to see more of them in the future. The Windows CE operating system is becoming increasingly mature as Microsoft keeps developing new versions, service packs, removing bugs and adding new features. Various companies are making IP telephones, home entertainment devices, handheld computers and other different gadgets based on the Windows CE operating system.

**7. Advantages and disadvantages**

After taking into account all relevant features of mobile applications it is possible to compare four mobile technologies described in this article on the basis of spread, capability, development complexity and the ability to produce financial profit for application developers. Each category falls in one of three ranges: Low, Medium and High. The result is shown in Table 6. Each of the four technologies has some comparable advantages but also some drawbacks so each technology has its own unique trade-offs. SMS applications are, simply put, the most attractive type of mobile applications. They represent a special breed – with the highest spread but with the least capabilities provided to the users. SMS applications will bring most profit to the developer but also, since they are typically n-tier client/server systems with heterogeneous technologies and highly dependant on Telecom's infrastructure, SMS applications will be complicated to develop. WAP applications have almost the same spread as SMS but their capabilities are a bit poor. They are severely constrained by mobile devices screen size. It is just not possible to display much data on a small screen. On the other hand, development of WAP applications is easy – like development of simple static or dynamic HTML pages. However, taking all into account, it is fair to say that WAP applications have not lived up to expectations and they never will. In all probability, newer technologies like Web browsers for PDAs, with special rendering of HTML and XHTML pages suitable for smaller screens, will replace WAP browsers. Advantages and disadvantages of J2ME and Windows CE mobile applications are almost the same but the most crucial difference is that Windows CE is used on PDAs with touch-screens and J2ME is for mobile phones with small screens and keyboards. Both technologies are very capable and have powerful applicative features but Windows CE is the most capable platform.

|  | SMS | WAP | J2ME | Windows CE |
|---|---|---|---|---|
| Spread | High | High | High/ Medium | Medium/ Low |
| Capability | Low | Medium/ Low | High/ Medium | High |
| Development complexity | High | Medium/ Low | Medium/ Low | Medium/ Low |
| Profit | High | Low | Medium/ Low | Medium/ Low |

Table 6 Comparative analysis of different mobile technologies

Windows CE is a real-time operating system built around a comprehensive software framework and it has rich user interface together with high screen resolution. On the other hand, since mobile phones are more common than mobile computers, J2ME is more widespread than Windows CE. Perhaps in the future, when more PDAs with voice call capabilities hit the market, this gap will be reduced.

## 8. Conclusion

It is impossible to give a straight yes or no answer to the question which mobile technology is simply the best. As can be seen in Table 6, each mobile technology - SMS, WAP, J2ME and Windows CE - has its own comparable advantages. None of the technologies can be totally ruled out. In a given situation it might be necessary to develop applications in any or even several of these four technologies. Project managers and software designers will always have to select the best mobile technology for a given set of software requirements but, simply put, there are just two appealing choices: SMS or J2ME/Windows CE. SMS is the best choice for a mobile application that can have a simple user interface, has to perform one specific task and has to reach the widest possible range of users. SMS applications do represent the biggest development challenge but they are financially the most lucrative solution for any developer of mobile applications. J2ME, especially with MIDP 2.0 standard, is the best solution for capable applications on mobile phones. Windows CE is in many ways the strongest mobile technology that is suited for mobile computers. However, the most important thing to notice is that the penetration of mobile communications worldwide is already high and, more importantly, it is steadily rising. Therefore, the number of potential users of mobile applications is constantly increasing as well and this can only be good news for software developers.

## 9. References

[1] R. Kalden, I. Meirick, and M. Meyer, "Wireless Internet Access based on GPRS", http://www.stephan-baucke.de/publications/mme/PCM_4_2000.pdf [2000]

[2] H. Schotten, "Evolution of 3G radio access techniques", in Proc. of the International Symposium '3G Infrastructure and Services' 3GIS, Athens, pp. 161-165, 2001.

[3] Y. Guo and H. Chaskar, "Class-based quality of service over air interfaces in 4G mobile networks", IEEE Commun. Mag., pages 132-137, March 2002.

[4] B.G. Evans and K. Baughan, "Visions of 4G", Electronics & Communication Engineering Journal, Vol. 12, No. 6, pp. 293-303, Dec. 2000.

[5] IRDA. IrDA Object Exchange Protocol (IrOBEX), January 1997.

[6] Infrared Data Association. "Technical Summary of IrDA DATA and IrDA CONTROL.", http://www.irda.org/standards/standards.asp [1999]

[7] IrDA SIR Data Specification. At http://www.irda.org/ standards/pubs/IrData.zip, February 1999.

[8] J Haartsen, "The Bluetooth Radio System", IEEE Personal Communications, pp. 28-36, Feb. 2000.

[9] Specification of the Bluetooth system, http://www.bluetooth.com [Dec. 1999]

[10] A. Baer. "The Wi-Fi boom; out and about, and online", The New York Times, December 12, 2002.

[11] J. Lansford, A. Stephens, and R. Nevo, "Wi-Fi (802.11b) and Bluetooth: Enabling coexistences", EEE Network, Vol. 15, pp. 20 27, September/October 2001.

[12] I. Wijegunaratne, M. Socic, and C. Chow, "An Architecture for Client/Server Application Software", Australian Computer Journal, Vol. 26, No. 2, pp. 30-41, 1994.

[13] S. Pehrson, "WAP - The catalyst of the mobile Internet", Ericsson Review, No. 1, pp. 14-19, 2000.

[14] L. Passani, "Creating WAP Services", Dr. Dobb's Journal of Software Tools, Vol. 25, No. 7, pp. 70, 73--75, 78, Jul. 2000.

[15] A. Schmidt, H. Schroder and F. Frick, "WAP: designing for small user interfaces", Proceedings of ACM CHI 2000 Conference on Human Factors in Computing Systems, Interactive posters, Vol. 2, pp. 187-188, 2000.

[16] Inc. Sun Microsystem, Mobile Information Device Profile (MIDP) Specification. http://java.sun.com/products/midp/ [2000]

[17] L. Aarnio, "Small-Scale Java Virtual Machines", http://www.cs.helsinki.fi/u/campa/teaching/j2me/papers/Small.pdf

[18] C. Muench, "The Windows CE Technology Tutorial: Windows Powered Solutions for the Developer". Addison Wesley Longman: Boston, 2000.

[19] A. H. James and Y. Wilson, "Building Powerful Platforms with Windows CE". Addison Wesley Professional; ISBN: 02016.